\documentclass{article}

\usepackage{microtype}
\usepackage{graphicx}
\usepackage{subfigure}
\usepackage{booktabs} 
\usepackage{hyperref}
\usepackage{hyperref}

\usepackage[accepted]{icml2025}

\usepackage{amsmath}
\usepackage{amssymb}
\usepackage{bm}
\newcommand{\bfx}{\bm{x}}
\usepackage{mathtools}
\usepackage{amsthm}
\newcommand{\gD}{\mathcal{D}}
\newcommand{\rvp}{\mathbf{p}}
\newcommand{\rvm}{\mathbf{m}}
\usepackage[capitalize,noabbrev]{cleveref}

\theoremstyle{plain}

\theoremstyle{definition}

\theoremstyle{remark}

\usepackage[textsize=tiny]{todonotes}

\icmltitlerunning{MolFORM: Multi-modal Flow Matching for Structure-Based Drug Design}

\begin{document}

\twocolumn[
\icmltitle{MolFORM: Multi-modal Flow Matching for Structure-Based Drug Design}



\icmlsetsymbol{equal}{*}

\begin{icmlauthorlist}
\icmlauthor{Jie Huang}{equal,osu}
\icmlauthor{Daiheng Zhang}{equal,rutgers}

\end{icmlauthorlist}

\icmlaffiliation{rutgers}{Department of Electrical and Computer Engineering, Rutgers University, New Brunswick, New Jersey, USA}
\icmlaffiliation{osu}{Department of Chemistry and Biochemistry, The Ohio State University, Columbus, Ohio, USA}

\icmlcorrespondingauthor{Daiheng Zhang}{dz367@rutgers.edu}

\icmlkeywords{Machine Learning, ICML}

\vskip 0.3in
]



\printAffiliationsAndNotice{\icmlEqualContribution} 

\begin{abstract}
Structure-based drug design (SBDD) seeks to generate molecules that bind effectively to protein targets by leveraging their 3D structural information. While diffusion-based generative models have become the predominant approach for SBDD, alternative non-autoregressive frameworks remain relatively underexplored. In this work, we introduce MolFORM, a novel generative framework that jointly models discrete (atom types) and continuous (3D coordinates) molecular modalities using multi-flow matching. To further enhance generation quality, we incorporate a preference-guided fine-tuning stage based on \textit{Direct Preference Optimization} (DPO), using Vina score as a reward signal. We propose a multi-modal flow DPO co-modeling strategy that simultaneously aligns discrete and continuous modalities, leading to consistent improvements across multiple evaluation metrics. The code is provided at: \url{https://github.com/huang3170/MolForm}.
\end{abstract}

\section{Introduction}
\label{Introduction}
Structure-based drug design (SBDD) \cite{anderson2003process} accelerates drug discovery by utilizing the three-dimensional structures of biological targets, enabling the efficient and rational design of molecules within a defined chemical space. Generative models have recently emerged as a powerful approach for streamlining the SBDD process by directly proposing candidate molecules, thus bypassing the need for exhaustive exploration of large chemical libraries. Advances in this area can be broadly categorized into two directions: autoregressive models ~\cite{luo20213d}, which formulate molecule generation as a sequential prediction task, and diffusion models ~\cite{guan20233d,guan2024decompdiff}, which draw inspiration from the iterative refinement process commonly used in image generation.
Despite the variety of non-autoregressive generative models, diffusion-based approaches have become the dominant paradigm. In SBDD, many extensions have been developed on top of diffusion models to better handle protein-ligand interactions, with a particular focus on improving binding affinity through task-specific objectives and interaction-aware designs ~\citep{huang2023protein,guan2024decompdiff}. In parallel, there has been growing interest in exploring alternative non-autoregressive frameworks such as Bayesian Flow Networks ~\cite{qu2024molcraft}, which have also demonstrated promising results, achieving state-of-the-art performance ~\cite{lin2024cbgbench} on several benchmark SBDD tasks.

In recent years, flow matching \cite{liu2022flow,lipman2022flow} has emerged as a widely studied generative modeling framework. Although flow matching is theoretically equivalent to diffusion models under certain conditions \cite{gao2025diffusionmeetsflow}, empirical performance can vary significantly depending on the choice of scheduling strategy. In image generation tasks, flow matching has been successfully scaled to large datasets and has demonstrated strong performance \cite{esser2024scaling,liu2023instaflow}. In the domain of AI for science, especially for molecular and protein generation tasks, researchers have also begun to explore the applicability of flow matching \cite{jing2024alphafold,geffner2025proteina, campbell2024generative}. Conceptually, flow matching offers a transport mapping interpretation from the perspective of ordinary differential equations (ODEs), providing a flexible modeling framework that enables task-specific adaptations. For structure-based drug design (SBDD), the generative task involves predicting both atom types and their 3D positions, which can be viewed as a combination of discrete and continuous modalities. Motivated by recent advances, we propose \textbf{MolFORM}, a novel framework for \textbf{Mol}ecular multi-modal \textbf{F}low-\textbf{O}ptimized \textbf{R}epresentation \textbf{M}atching. Our results show that carefully designed flow-based models can achieve performance comparable to SOTA diffusion-based approaches.

Furthermore, we incorporate a preference-guided fine-tuning stage using \textit{Direct Preference Optimization} (DPO), leveraging the Vina score as a chemically informed reward signal to refine the base model. We first propose a discrete flow matching variant of DPO under a uniform noising scheme, which enables preference alignment in the atom type space. More importantly, we demonstrate that the full benefit arises from a multi-flow DPO co-modeling strategy that jointly aligns preferences over both discrete atom types and continuous 3D positions. This additional step leads to notable improvements in the quality of generated molecules, with our experiments showing that the gains from DPO fine-tuning surpass those of standard diffusion-based models. To the best of our knowledge, this is the first work to apply preference alignment in a multi-flow setting, highlighting its potential for broader adoption in similar generative tasks.

\section{Related work}
\subsection{Structure-Based Drug Design.}
With the increasing availability of structural data, generative models have attracted significant attention for structure-based molecule generation. Early methods~\citep{skalic2019target} utilized sequence generative models to produce SMILES representations from protein contexts. Driven by advancements in 3D geometric modeling, subsequent studies directly generate molecules in 3D space. For instance, \citet{ragoza2022generating} employ voxelized atomic density grids within a Variational Autoencoder framework. Other methods use autoregressive models to sequentially place atoms or chemical groups~\citep{luo20213d, peng2022pocket2mol}, while FLAG~\citep{zhang2023molecule} and DrugGPS~\citep{zhang2023learning} leverage chemical priors to construct realistic ligand fragments incrementally. More recently, diffusion models have demonstrated notable success by progressively denoising atom types and coordinates, maintaining SE(3)-equivariant symmetries~\citep{guan20233d, lin2022diffbp, schneuing2022structure, huang2023protein, guan2024decompdiff, zhang2024rectified}. Despite these advances, existing models often struggle with generating molecules simultaneously optimized for multiple desirable properties, such as binding affinity, synthesizability, and low toxicity—critical considerations in drug discovery~\citep{d2012multi}.

\subsection{Flow Matching.}
Flow matching~\citep{lipman2022flow,liu2022flow} is a continuous-time generative modeling framework that generalizes diffusion models by learning a vector field to transport a simple prior $q(\bfx)$ toward the data distribution $p_{\text{data}}(\bfx)$. It defines a conditional path $p_t(\bfx \mid \bfx_1)$ that interpolates between $q(\bfx)$ and the target $\delta(\bfx - \bfx_1)$, and learns the marginal vector field $v(\bfx, t) = \mathbb{E}_{\bfx_1 \sim p_t(\bfx_1 \mid \bfx)}[u_t(\bfx \mid \bfx_1)]$ using a neural network ${v}_\theta(\bfx, t)$. Diffusion models can be seen as a special case under Gaussian interpolation~\citep{gao2025diffusionmeetsflow}. Flow matching has demonstrated strong performance in image and video generation~\citep{liu2023instaflow,esser2024scaling}, and is gaining traction in scientific domains such as protein generation ~\citep{campbell2024generative}, protein conformation generation~\citep{jing2024alphafold}, where it naturally handles both discrete atom types and continuous 3D coordinates.

\subsection{Preference alignment of generative models.} 
While maximizing data likelihood is standard in generative modeling, it often fails to align with downstream user preferences. Reinforcement learning from human feedback (RLHF)~\citep{ziegler2020finetuning, ouyang2022training} and its variants have been widely adopted to align large language models with human intent. Recent efforts extend these ideas to diffusion models, treating generation as a multi-step decision process~\citep{uehara2024feedback, wallace2023diffusion}. Direct Preference Optimization (DPO)~\citep{rafailov2023direct} offers a simpler alternative by bypassing reinforcement learning and directly optimizing models against pairwise preference data. DPO has shown competitive results in both language and image domains~\citep{wallace2023diffusion, zhou2024antigen}. Meanwhile, we also observe that some studies~\citep{gu2024aligning, cheng2024decomposed} have begun to incorporate preference alignment into structure-based drug design (SBDD), highlighting its potential to improve biological plausibility and design success. However, applying DPO to jointly model discrete and continuous variables remains largely unexplored.

\begin{figure*}
  \centering
  \includegraphics[width=0.95\textwidth]{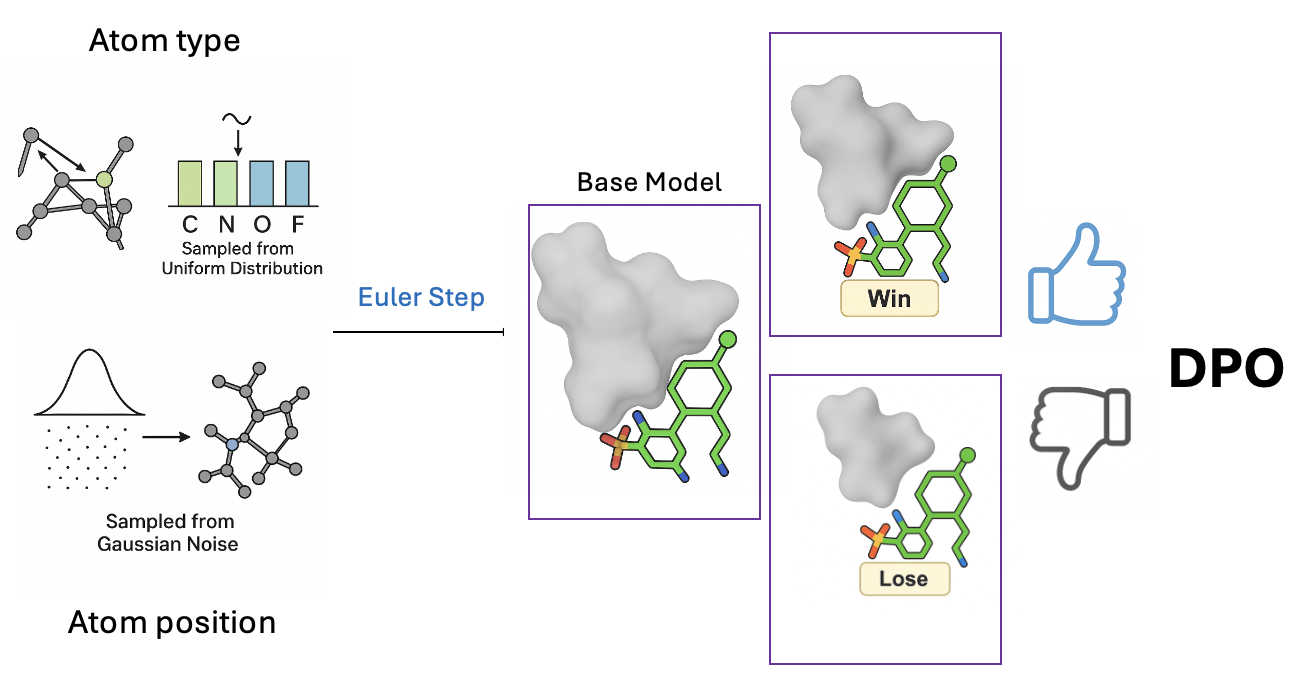}
  \caption{Overview of MolFORM. This workflow can be summarized as two steps: 1) Employs multi-flow generation to construct the base model. 2) Applies DPO to fine-tune the dual modalities, using the Vina score as the reward.}
  \label{fig:Model}
\end{figure*}

\section{Methods}

\subsection{Problem definitions.}
We aim to generate ligand molecules that are capable of binding to specific protein binding sites, by modeling \( p(M|P) \). 
We represent the protein pocket as a collection of \( N_P \) atoms, \( P = \{ (x^{i}_P, v^{i}_P) \}_{i=1}^{N_P} \). Similarly, the ligand molecule can be represented as a collection of \( N_M \) atoms, \( M = \{ (x^{i}_M, v^{i}_M) \}_{i=1}^{N_M} \), where \( x^i_M \in \mathbb{R}^3 \) represents atom position and \( v^{i}_M \in [k] \)  the $k$ possible atom types. 
The number of atoms \( N_M \) can be sampled from an empirical distribution \citep{hoogeboom2022equivariant,guan20233d}. For brevity, the ligand molecule is denoted as \(\ M = \{ \mathbf{X}, \mathbf{V} \} \) where  \( \mathbf{X} \in \mathbb{R}^{N_M \times 3} \) and \( \mathbf{V} \in [k]^{N_M \times {K}} \).

\subsection{Multi-modal Flow matching}
\label{sec:cfm}

The Conditional Flow Matching (CFM) framework~\cite{lipman2022flow,liu2022flow} learns a time-dependent flow $\psi_t: [0,1] \times \mathbb{R}^d \rightarrow \mathbb{R}^d$ that transports samples from a source distribution $p_0$ to a target distribution $p_1$, governed by the ODE $\frac{d}{dt}\mathbf{x}_t = u_t(\mathbf{x}_t)$ with $\mathbf{x}_t = \psi_t(\mathbf{x}_0)$. Since the exact vector field $u_t$ is generally intractable, the conditional formulation enables a tractable objective by conditioning on both source and target samples, where the $ v_\theta$ is a time-
depended, learnable model with parameters $\theta$:
\begin{equation}
    \mathcal{L}_{\text{CFM}}(\theta) = \mathbb{E}_{t, \mathbf{x}_0, \mathbf{x}_1} \left\| v_\theta(\mathbf{x}_t, t) - u_t(\mathbf{x}_t | \mathbf{x}_0, \mathbf{x}_1) \right\|^2.
\end{equation}
In our structure-based drug design (SBDD) task, we model each atom type as a categorical variable, where $x \in \mathcal{K}$ and $\mathcal{K} = \{1, \dots, K\}$ is the set of possible atom types with cardinality $K$. To model the generative process over atom types, we adopt the Discrete Flow Matching (DFM) framework \cite{campbell2024generative}, which constructs a time-dependent probability flow $p_t(v)$ that evolves from a known noise distribution $p_0(v)$ to the empirical data distribution $p_1(v) = p_{\mathrm{data}}(v)$ over a time interval $t \in [0, 1]$.

This flow is defined via a family of conditional distributions $\pi_t(v_t \mid v_1)$, referred to as conditional flows, where $v_1 \in [k]$ is the true atom type and $v_t$ is its corrupted version at time $t$. We adopt the uniform corruption process:
\begin{equation}
\pi_t(v_t \mid v_1) = (1 - t) \cdot \mathrm{Uniform}([k]) + t \cdot \delta_{v_1}(v_t),
\end{equation}
where $\delta_{v_1}(v_t)$ is the Kronecker delta function (equal to 1 if $v_t = v_1$, and 0 otherwise), and $\mathrm{Uniform}([k]) = 1/k$ for all $v_t \in [k]$. This defines a simple interpolation between a uniform noise distribution $p_0(v) = \mathrm{Uniform}([k])$ at $t=0$ and the data distribution $p_1(v) = p_{\mathrm{data}}(v)$ at $t=1$.
The marginal distribution at time $t$ is given by averaging over the data distribution:
\begin{equation}
p_t(v_t) = \mathbb{E}_{v_1 \sim p_{\mathrm{data}}} \left[ \pi_t(v_t \mid v_1) \right].
\end{equation}

To train the model, we learn a denoising distribution $p^{\theta}(v^i_1 \mid \mathbf{v}_t)$ that predicts the original atom type $v^i_1 \in [k]$ from a corrupted ligand type sequence $\mathbf{v}_t = \{v^i_t\}_{i=1}^{N_M}$. The model is optimized using a standard cross-entropy loss:

\begin{equation}
    \mathcal{L}_{\mathrm{CE}} = \mathbb{E}_{t,\, \mathbf{v}_1,\, \mathbf{v}_t \sim \pi_t(\cdot \mid \mathbf{v}_1)} \left[ - \sum_{i=1}^{N_M} \log p^{\theta}(v^i_1 \mid \mathbf{v}_t) \right],
\end{equation}

where $\pi_t(v^i_t \mid v^i_1)$ denotes the uniform corruption process.

To improve training stability, especially when $t \to 1$, we adopt a reparameterized formulation inspired by diffusion-based objectives. Instead of regressing the velocity field directly, we predict the target sample $\hat{\mathbf{x}}_1 = v_\theta(\mathbf{x}_t, t)$ and define the loss in terms of reconstructed flows:
\begin{equation}
    \mathcal{L}_{\text{reparam}}(\theta) = \mathbb{E}_{t, \mathbf{x}_0, \mathbf{x}_1} \left\| u_t(\mathbf{x}_t | \hat{\mathbf{x}}_1, \mathbf{x}_0) - u_t(\mathbf{x}_t | \mathbf{x}_1, \mathbf{x}_0) \right\|^2.
\end{equation}
On Euclidean manifolds, where $u_t(\mathbf{x}_t | \mathbf{x}_1, \mathbf{x}_0) = \mathbf{x}_1 - \mathbf{x}_0$, this reduces to a simple mean squared error:
\begin{equation}
    \mathcal{L}_\text{pos} = \mathbb{E}_{t, \mathbf{x}_0, \mathbf{x}_1} \left\| \hat{\mathbf{x}}_1 - \mathbf{x}_1 \right\|_2^2.
\end{equation}
Similarly, we reparameterized the cross-entropy loss to predict the clean sample $\hat{\mathbf{v}}_1$ and match the conditional flows induced by $\hat{\mathbf{v}}_1$ and $\mathbf{v}_1$. Here the loss function is expressed as:
\begin{equation}
    \mathcal{L}_\text{type} = \mathbb{E}_{t,\, \mathbf{v}_0, \mathbf{v}_1} \left[ \mathrm{CE} \left( \hat{\mathbf{v}}_1,\, \mathbf{v}_1 \right) \right],
\end{equation}
This reparameterization has been shown to enhance numerical stability, especially when sampling with large $t$.

\textbf{\begin{figure*}
  \centering
  \includegraphics[width=\textwidth]{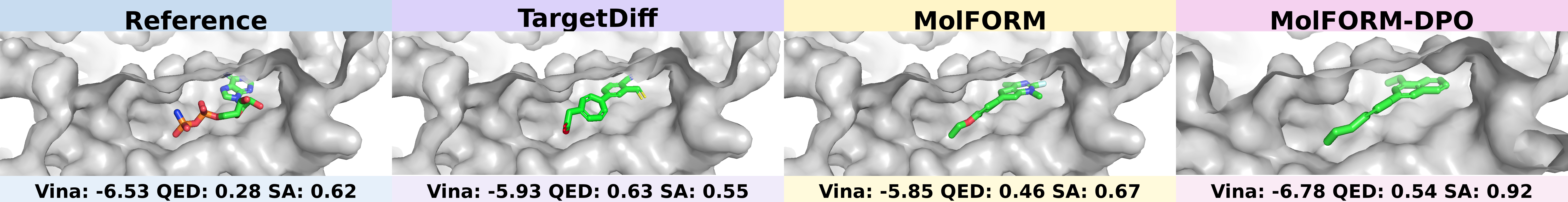}
  \caption{Visualizations of reference molecules and generated ligands for protein pockets (4yhj) generated by Reference, Targetdiff, MolFORM and MolFORM-DPO. Vina score, QED, and SA are reported below.}
  \label{fig:visul}
\end{figure*}}

\paragraph{Chamfer loss}  
To promote accurate geometric alignment between predicted and ground-truth molecular structures, we incorporate a Chamfer loss defined over atomic point clouds. Given two point sets $\hat{\mathbf{x}}_1 = \{\hat{x}_i\}_{i=1}^N$ and $\mathbf{x}_1 = \{x_j\}_{j=1}^M$ representing predicted and reference atomic positions respectively, the Chamfer distance is defined as:
\begin{align}
\label{Chamfer}
\mathcal{L}_{\text{Chamfer}} =\ & \frac{1}{N} \sum_{\hat{x} \in \hat{\mathbf{x}}_1} \min_{x \in \mathbf{x}_1} \|\hat{x} - x\|_2 \nonumber \\
& + \frac{1}{M} \sum_{x \in \mathbf{x}_1} \min_{\hat{x} \in \hat{\mathbf{x}}_1} \|x - \hat{x}\|_2.
\end{align}
This bidirectional distance encourages each predicted atom to be close to some reference atom, and vice versa, effectively guiding the flow matching process to learn a distribution that preserves spatial fidelity to the ground-truth.



The final loss is shown below:
\begin{equation}
\mathcal{L} = \mathcal{L}_\text{pos} + \mathcal{L}_\text{type} + \lambda \cdot \mathcal{L}_{\text{Chamfer}}
\label{eq:3loss}
\end{equation}
where $\lambda$ is a weighting hyperparameter that balances the contribution of the Chamfer loss.

\begin{table*}[!t]
\centering
\scalebox{0.68}{
\begin{tabular}{c|c|c|c|c|cc|cc|cc|cc}
\hline
Model & 
\begin{tabular}{c}Vina Score  ($\downarrow$) \\ \end{tabular} & 
\begin{tabular}{c}Vina Min ($\downarrow$) \end{tabular} & 
\begin{tabular}{c}Vina Dock ($\downarrow$) \end{tabular} & 
\begin{tabular}{c}Diversity ($\uparrow$) \end{tabular} & 
\multicolumn{2}{|c|}{QED ($\uparrow$)} & 
\multicolumn{2}{|c|}{SA ($\uparrow$)} & 
\multicolumn{2}{|c|}{Static Geometry ($\downarrow$)} & 
\multicolumn{2}{|c}{Clash ($\downarrow$)} \\

 & Avg.& Avg.& Avg.&Avg. & Avg. & Med. & Avg. & Med. & JSD$_{\mathrm{BL}}$ & JSD$_{\mathrm{BA}}$ & Ratio$_{\mathrm{cca}}$ & Ratio$_{\mathrm{cm}}$ \\
\hline
LiGAN        & \textbf{-6.47} & \underline{-7.14} & \underline{-7.70} &0.66& 0.46 & 0.46 & \textbf{0.66} & \textbf{0.66} & 0.4645 & 0.5673 & \textbf{0.0096} & \textbf{0.0718}  \\
3DSBDD       & -     & -3.75 & -6.45 & 0.70&0.48 & 0.48 & 0.63 & 0.63 & 0.5024 & 0.3904 & 0.2482 & 0.8683  \\
GraphBP      & -     & -     & -4.57 & \textbf{0.79}&0.44 & 0.44 & 0.64 & 0.64 & 0.5182 & 0.5645 & 0.8634 & 0.9974  \\
Pocket2Mol   & -5.23 & -6.03 & -7.05 & 0.69&0.39 & 0.39 & \underline{0.65} & \underline{0.65} & 0.5433 & 0.4922 & 0.0576 & 0.4499  \\
TargetDiff   & -5.71 & -6.43 & -7.41 & 0.72&0.49 & 0.49 & 0.60 & 0.60 & 0.2659 & 0.3769 & 0.0483 & 0.4920  \\
DiffSBDD     & -     & -2.15 & -5.53 & -&0.49 & 0.49 & 0.34 & 0.34 & 0.3501 & 0.4588 & 0.1083 & 0.6578 \\
DiffBP       & -     & -     & -7.34 & - &0.47 & 0.47 & 0.59 & 0.59 & 0.3453 & 0.4621 & 0.0449 & 0.4077  \\
FLAG         & -     & -     & -3.65 & -&0.41 & 0.41 & 0.58 & 0.58 & 0.4215 & 0.4304 & 0.6777 & 0.9769  \\
D3FG         & -     & -2.59 & -6.78 & -& 0.49 & 0.49 & \textbf{0.66} & \textbf{0.66} & 0.3727 & 0.4700 & 0.2115 & 0.8571  \\
DecompDiff   & -5.18 & -6.04 & -7.10 & 0.68&0.49 & 0.49 & \textbf{0.66} & \textbf{0.66} & \underline{0.2576} & \underline{0.3473} & 0.0462 & 0.5248   \\
MolCraft     & -6.15 & -6.99 & -7.79 & 0.72& 0.48 & 0.48 & \textbf{0.66} & \textbf{0.66} & \textbf{0.2250} & \textbf{0.2683} & 0.0264 & 0.2691 \\
VoxBind      & \underline{-6.16} & -6.82 & -7.68 & -& \textbf{0.54} & \textbf{0.54} & 0.65 & \underline{0.65} & 0.2701 & 0.3771 & \underline{0.0103} & \underline{0.1890} \\
MolFORM      & -5.42 & -6.42 & -7.50 & \underline{0.78}& 0.48 & 0.49 & 0.60 & 0.58 &  0.3225 & 0.5535 & 0.0310 & 0.4474\\
MolFORM-DPO & \underline{-6.16} & \textbf{-7.18} & \textbf{-8.13} & 0.77& \underline{0.50} &  \underline{0.51} & \underline{0.65} &  0.63 &  0.3215 & 0.5584 & 0.0188 & 0.2525 \\
\midrule
Reference    & -6.36 & -6.71 & -7.45 & -&0.48 & 0.47 & 0.73 & 0.74 &  - & - & - & -  \\
\bottomrule
\end{tabular}
}
\vspace{10pt}
\caption{Combined results for binding affinity, chemical properties, and geometry/clash metrics. (↑)/(↓) denote better. Top 2 results are marked in bold and underlined. The result from baseline model is quoted from ~\citet{lin2024cbgbench}.}
\label{tab:combined_metrics}
\end{table*}

\subsection{Sampling}
\label{sec:sec:sampling}
We consider two types of generative sampling procedures, corresponding to the discrete atom types and continuous atomic coordinates.

\paragraph{Continuous sampling (atom positions)}
For atomic coordinates, we simulate trajectories via the learned velocity field \( v_\theta(\mathbf{x}_t, t) \) using Euler integration in \( N \) steps:
\begin{equation}
\mathbf{x}_{t + \frac{1}{N}} = \mathbf{x}_t + \frac{1}{N} \cdot v_\theta(\mathbf{x}_t, t),\quad t \in \left\{0, \tfrac{1}{N}, \dots, \tfrac{N-1}{N} \right\}.
\end{equation}
Starting from \( \mathbf{x}_0 \sim p_0 \), where  $p_0$ is a Gaussian distribution, we integrate the dynamics to obtain the final conformation \( \mathbf{x}_1 \).

\paragraph{Discrete sampling (atom types)}
Given an initial sample \( v_0 \sim p_0 \), where $p_0$ is a  discrete uniform distribution over the number of atom types, we simulate the forward trajectory using Euler integration:
\begin{equation}
v_{t + \Delta t} \sim \mathrm{Cat}\left( \delta_{v_t} + R_t(v_t, \cdot)\, \Delta t \right).
\end{equation}
The rate matrix can be computed as an expectation over conditional flows:
\begin{equation}
R_t(v_t, j) = \mathbb{E}_{v_1 \sim q(v_1 \mid v_t)} \left[ R_t(v_t, j \mid v_1) \right],
\end{equation}
where the posterior is given by
\begin{equation}
q(v_1 \mid v_t) = \frac{\pi_t(v_t \mid v_1) \cdot p_{\mathrm{data}}(v_1)}{p_t(v_t)}.
\end{equation}

\subsection{Direct Preference Optimization(DPO) on Multi-Flow}

Building upon these developments, we introduce a Multi-flow DPO approach that jointly optimizes both continuous and discrete data using Reinforcement Learning from Human Feedback (RLHF).

Given winning samples $x_1^w$ and losing samples $x_1^l$ for a given sample $x_1$ under a preference $\mathcal{P}$, we construct a dataset of preference pairs \( \mathcal{D} = \{ (p, x_1^w, x_1^l) \} \), where $p$ denotes the protein condition. Using the loss function in Equation~\eqref{eq:DPOloss}, we bypass the need to explicitly estimate a reward function and instead directly optimize the parameterized model $p_\theta$. Here, $\log\sigma$ denotes the log-sigmoid function, $\beta$ is a scaling (regularization) coefficient, $p_\theta$ is the learnable model, and $p_{\text{ref}}$ is a fixed reference model.
\begin{align}
L_{\mathrm{DPO}}(\theta) ={}& -\mathbb{E}_{(x_1^w, x_1^l) \sim \mathcal{D}, {p^{\theta}_{1 | t}(x_1 | x_t)}, t \sim \mathcal{U}[0,1]} \Big[ \log \sigma  \notag \\
&\quad \Big(\beta \log \frac{p_\theta(x_1^w \mid p)}{p_{\text{ref}}(x_1^w \mid p)}  - \beta \log \frac{p_\theta(x_1^l \mid p)}{p_{\text{ref}}(x_1^l \mid p)} 
\Big) \Big]
\label{eq:DPOloss}
\end{align}

The DPO loss can be decomposed into three parts \ref{eq:posDPOLoss}, \ref{eq:pointcloudDPOLoss} and \ref{eq:atomtypeDPOLoss} as below, aligning with Equation~\eqref{eq:3loss}, to reflect discrete and continuous components:
\begin{flalign}
L_{\mathrm{DPO}}^x(\theta) ={}& 
-\mathbb{E}_{(x_1^w, x_1^l) \sim \mathcal{D},\; t \sim \mathcal{U}[0,1],   {p^{\theta}_{1 | t}(x_1 | x_t)}} \notag \\
&\quad
\Big[ \log \sigma \Big(  -\beta \big( 
\| x_1^w - \hat{x}_{1,\theta}^w \|^2 
- \| x_1^w - \hat{x}_{1,\text{ref}}^w \|^2 \notag \\
&\quad 
- \| x_1^l - \hat{x}_{1,\theta}^l \|^2 
+ \| x_1^l - \hat{x}_{1,\text{ref}}^l \|^2 
\big) \Big) \Big]
\label{eq:posDPOLoss}
\end{flalign}

\begin{flalign}
& L_{\mathrm{DPO}}^{\text{point cloud}}(\theta) ={} 
-\mathbb{E}_{(x_1^w, x_1^l) \sim \mathcal{D},\; t \sim \mathcal{U}[0,1],   {p^{\theta}_{1 | t}(x_1 | x_t)}} \notag \\
&\quad  \Big[ 
\log \sigma \Big(-\beta \big( 
\mathcal{L}_{\text{Chamfer}}(x_1^w, \hat{x}_{1,\theta}^w) 
- \mathcal{L}_{\text{Chamfer}}(x_1^w, \hat{x}_{1,\text{ref}}^w) \notag \\
&\qquad 
- \mathcal{L}_{\text{Chamfer}}(x_1^l, \hat{x}_{1,\theta}^l) 
+ \mathcal{L}_{\text{Chamfer}}(x_1^l, \hat{x}_{1,\text{ref}}^l)
\big) \Big) \Big]
\label{eq:pointcloudDPOLoss}
\end{flalign}

\begin{align}
L_{\mathrm{DPO}}^{\text{v}}(\theta) ={}& 
-\mathbb{E}_{(x_1^w, x_1^l) \sim \mathcal{D},\; t \sim \mathcal{U}[0,1]} \Big[ \log \sigma \Big( \notag \\
& \quad -\beta \big( 
\mathcal{D}^{\theta}_\text{ref}(x_t^w|x_1^w) 
- \mathcal{D}_\text{ref}^{\theta}(x_t^l|x_1^l) 
\big) \Big) \Big]
\label{eq:atomtypeDPOLoss}
\end{align}

with
\begin{align}
\mathcal{D}^{\theta}_\text{ref}(x_t|x_1) ={}& 
\sum_{j \neq x_t} R_t^{q}(x_t, j | x_1) 
\log \frac{R_t^{\theta}(x_t, j)}{R_t^{\text{ref}}(x_t, j)} \notag \\
& + R_t^{\text{ref}}(x_t, j) 
- R_t^{\theta}(x_t, j) \: 
\label{eq: d formation}
\end{align}

The values $\hat{x}_{1,\theta}^l$ and $\hat{x}_{1,\theta}^w$ are the predictions from the model $p_\theta$, while $\hat{x}_{1,\text{ref}}^l$ and $\hat{x}_{1,\text{ref}}^w$ are the corresponding predictions from the reference model $p_{\text{ref}}$. 

In the discrete space, the rate matrix $R_t{(x_t, x_t+\Delta t)}$ is defined as the expectation of the conditional rate matrix as shown in the equation~\eqref{eq: rate matrix}.
\begin{align}
R_t(x_t, x_{t+\Delta t}) = \mathbb{E}_{p^{\theta}_{1 | t}(x_1 | x_t)} \left[ R^q_t(x_t, x_{t+\Delta t} | x_1) \right]
\label{eq: rate matrix}
\end{align}
By choosing the uniform noise initialization, $\mathrm{Uniform}([k])$, the unconditional rate matrix becomes equation~\eqref{eq:rate_matrix_process}\cite{campbell2024generative}. 

\begin{align}
    R_t(x_t, x_{t+\Delta t})  
    &= \frac{1}{1 - t} \, p^\theta_{1 \mid t}(x_1 = x_{t+\Delta t} \mid x_t),
    \label{eq:rate_matrix_process} \\
    & \text{for } x_{t+\Delta t} \neq x_t \nonumber
\end{align}

By substituting equation~\eqref{eq:rate_matrix_process} into equation~\eqref{eq: d formation} and equation~\eqref{eq:atomtypeDPOLoss}, we can get the final loss of DPO of the discrete space in equation~\eqref{eq: final_dpo_loss}.

\begin{align}
L_{\text{DPO}}^v(\theta) ={}& - \mathbb{E}_{(x_1^w, x_1^l) \sim \mathcal{D},\, t \sim \mathcal{U}[0,1],   {p^{\theta}_{1 | t}(x_1 | x_t)}} 
\left[ \log \sigma \left( - \frac{\beta}{1 - t} \right. \right. \notag \\
& \left. \left. \cdot \left( 
\log \frac{p_{1|t}^{\theta}(x_1^w \mid x_t^w)}{p_{1|t}^{\text{Ref}}(x_1^w \mid x_t^w)}
- \log \frac{p_{1|t}^{\theta}(x_1^l \mid x_t^l)}{p_{1|t}^{\text{Ref}}(x_1^l \mid x_t^l)}
\right) \right) \right]
\label{eq: final_dpo_loss}
\end{align}
Equation~\eqref{eq: final_dpo_loss} is the DPO formulation for discrete flow matching under uniform noising.

\subsection{Time Scheduling}
\paragraph{Training scheduling} We adopt a $t$-sampling strategy that emphasizes large time steps to improve the accuracy of local structural details. Following the scheduling method from \citet{geffner2025proteina}, we sample time as
\begin{equation}
    p(t) = 0.02\, \mathcal{U}(0,1) + 0.98\, \mathcal{B}(1.9, 1.0),
\end{equation}
where \( \mathcal{U} \) and \( \mathcal{B} \) denote the uniform and Beta distributions, respectively. This schedule reduces weight on early time steps (where global structure is more variable) and concentrates training on later steps to improve geometric fidelity in atom-level predictions.

\paragraph{Sampling scheduling}
In conventional flow matching, the sampling trajectory typically uses a uniform discretization of time \( t \in [0, 1] \) with \( N \) equally spaced steps. However, we observe that the model requires finer granularity near \( t = 1 \), where more detailed structure is synthesized. To better capture this behavior and improve generation quality, we adopt a non-uniform time discretization strategy when integrating the ODE using the Euler method. Specifically, we allocate 60 steps in the interval \( [0, 0.8] \) and 40 steps in \( [0.8, 1] \), allowing the model to focus more on the final refinement phase.

\section{Experiments}


\subsection{Experiment Setup}
\label{sec:Experimental_Setup}
\paragraph{Datasets}
Our experiments were conducted using the CrossDocked2020 dataset \citep{francoeur2020three}. Consistent with prior studies ~\citep[e.g.,][]{peng2022pocket2mol,guan20233d,guan2024decompdiff}, we adhered to the same dataset filtering and partitioning methodologies. We further refined the 22.5 million docked protein binding complexes, characterized by an RMSD $<  1 \AA $, and sequence identity less than 30\%. This resulted in a dataset comprising 100,000 protein-binding complexes for training, alongside a set of 100 novel complexes designated for testing.
\paragraph{DPO dataset}
Our data processing strategy for DPO follows the methodology proposed in ~\citet{gu2024aligning}. We preprocess the dataset into a preference format $\gD = \{(\rvp, \rvm^w, \rvm^l)\}$, where $\rvp$ denotes the protein pocket, $\rvm^w$ the preferred ligand, and $\rvm^l$ the less preferred one. For each pocket, we sample two candidate ligands and assign preference based on a user-defined reward, mainly binding energy (e.g., Vina score). Since the affinity labels are continuous, we follow the strategy in~\citet{gu2024aligning} and choose the molecule with the worst score as the dispreferred sample $\rvm^l$. This encourages a larger reward gap between $\rvm^w$ and $\rvm^l$, which has been shown to improve learning. 

\paragraph{Model architecture}
Inspired from recent progress in equivariant neural networks \citep{satorras2021n}, 
we model the interaction between the ligand molecule atoms and the protein atoms with a SE(3)-Equivariant GNN, the atom hidden embedding and coordinates are updated alternately in each layer, which follows \citet{guan20233d}. Our model architecture is plotted in Figure \ref{fig:Model}.

\begin{table}[!t]
\centering
\resizebox{0.47\textwidth}{!}{
\begin{tabular}{c|cccccc}
\toprule
Method & C-C & C-N & C-O & C=C & C=N & C=O \\
\midrule
LiGan      & 0.4986 & 0.4146 & 0.4560 & 0.4807 & 0.4776 & 0.4595 \\
3DSBDD     & 0.2090 & 0.4258 & 0.5478 & 0.5170 & 0.6701 & 0.6448 \\
GraphBP    & 0.5038 & 0.4231 & 0.4973 & 0.6235 & 0.4629 & 0.5986 \\
Pocket2Mol & 0.5667 & 0.5698 & 0.5433 & 0.4787 & 0.5989 & 0.5025 \\
TargetDiff & 0.3101 & 0.2490 & 0.3072 & 0.1715 & 0.1944 & 0.3629 \\
DiffSBDD   & 0.3841 & 0.3708 & 0.3291 & 0.3043 & 0.3473 & 0.3647 \\
DiffBP     & 0.5704 & 0.5256 & 0.5090 & 0.6161 & 0.6314 & 0.5296 \\
FLAG       & 0.3460 & 0.3770 & 0.4433 & 0.4872 & 0.4464 & 0.4292 \\
D3GF       & 0.4244 & 0.3227 & 0.3895 & 0.3860 & 0.3570 & 0.3566 \\
DecompDiff & 0.2562 & 0.2007 & 0.2361 & 0.2590 & 0.2844 & 0.3091 \\
MolCraft   & 0.2473 & 0.1732 & 0.2341 & 0.3040 & 0.1459 & 0.2250 \\
VoxBind    & 0.3335 & 0.2577 & 0.3507 & 0.1991 & 0.1459 & 0.3334 \\
MolFORM    & 0.4282 & 0.3419 & 0.3824 & 0.2898 & 0.3239 & 0.3583 \\
\bottomrule
\end{tabular}
}
\caption{JSD Bond Length Comparisons across different methods.}
\label{tab:jsd_bl}
\end{table}

\paragraph{Baselines.} 
We select all baseline models from CBGbench~\cite{lin2024cbgbench} for comparison with our method. Early structure-based drug design (SBDD) methods are built on voxel grids with deep neural networks, such as LiGAN~\citep{ragoza2022generating}, which generates atom voxelized density maps using variational autoencoders (VAE) and convolutional neural networks (CNNs), and 3DSBDD~\citep{3dsbdd}, which predicts atom types on grids with graph neural networks (GNNs) in an auto-regressive manner. The development of equivariant graph neural networks (EGNNs) enables direct generation of 3D atom positions, as seen in Pocket2Mol~\citep{peng2022pocket2mol} and GraphBP~\citep{liu2022generating}, which use auto-regressive strategies with normalizing flows. Diffusion-based methods such as TargetDiff~\citep{guan20233d}, DiffBP~\citep{lin2022diffbp}, and DiffSBDD~\citep{schneuing2024structure} generate atom types and positions using denoising diffusion probabilistic models. Recent methods incorporate domain knowledge to guide generation: FLAG~\citep{zhang2023molecule} and D3FG~\citep{lin2023functional} use fragment motifs for coarse molecular generation; DecompDiff~\citep{guan2024decompdiff} uses scaffold and arm clustering with Gaussian process models for atom positions. More recent advances like MolCraft~\citep{qu2024molcraft} and VoxBind~\cite{pinheiro2024structure} apply new generative modeling strategies, including Bayesian flow networks and voxel-based diffusion with walk-jump sampling.

\begin{figure}[t]
  \centering
  \includegraphics[width=0.48\textwidth]{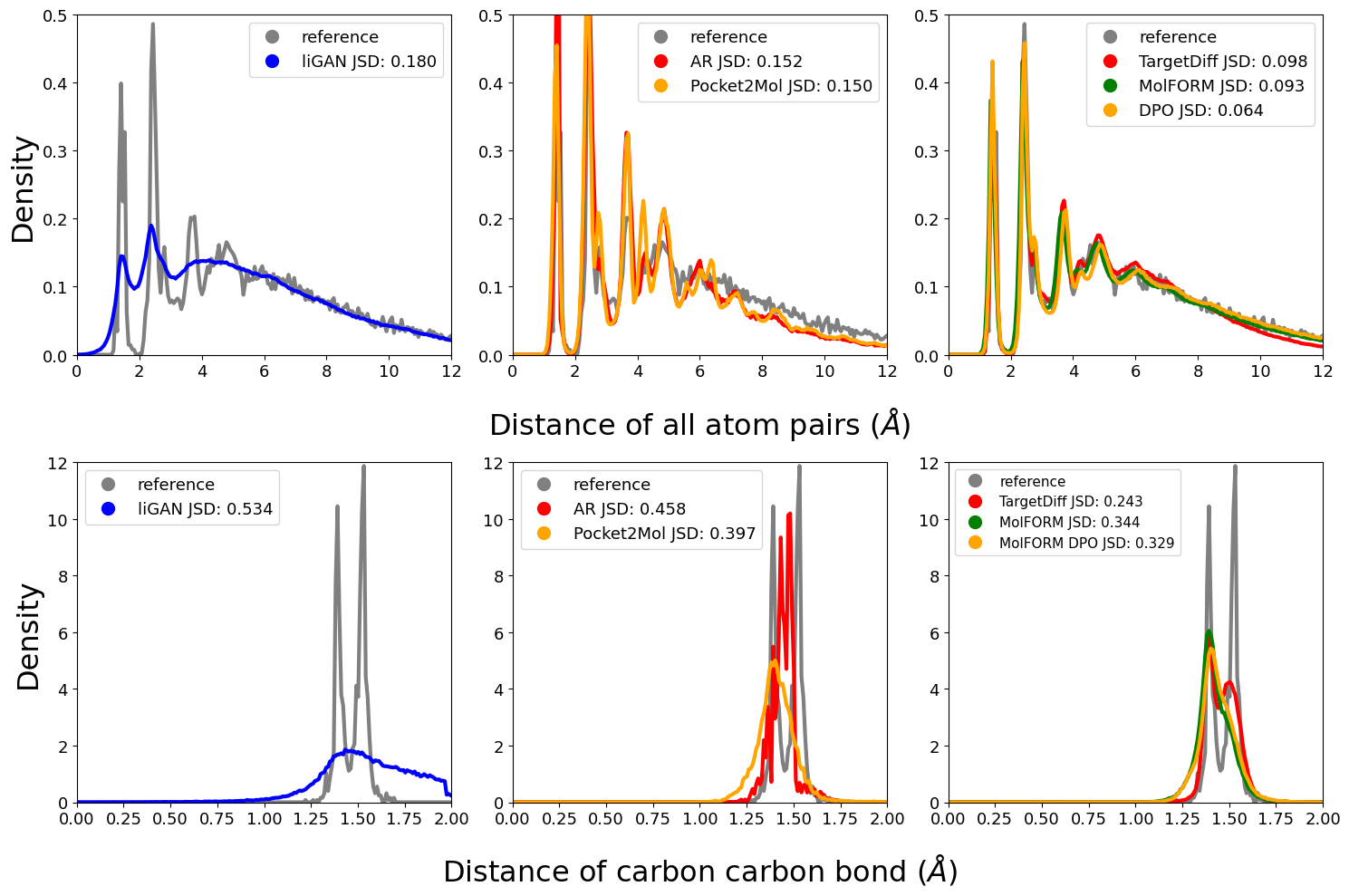}
  \caption{Comparing the distribution for distances of allatom (top row) and carbon-carbon pairs (bottom row) for reference molecules in the test set (gray) and model generated molecules (color).}
  \label{fig:dist}
\end{figure}

\subsection{Experiment result}

\begin{figure}[h]
  \centering
  \includegraphics[width=0.48\textwidth]{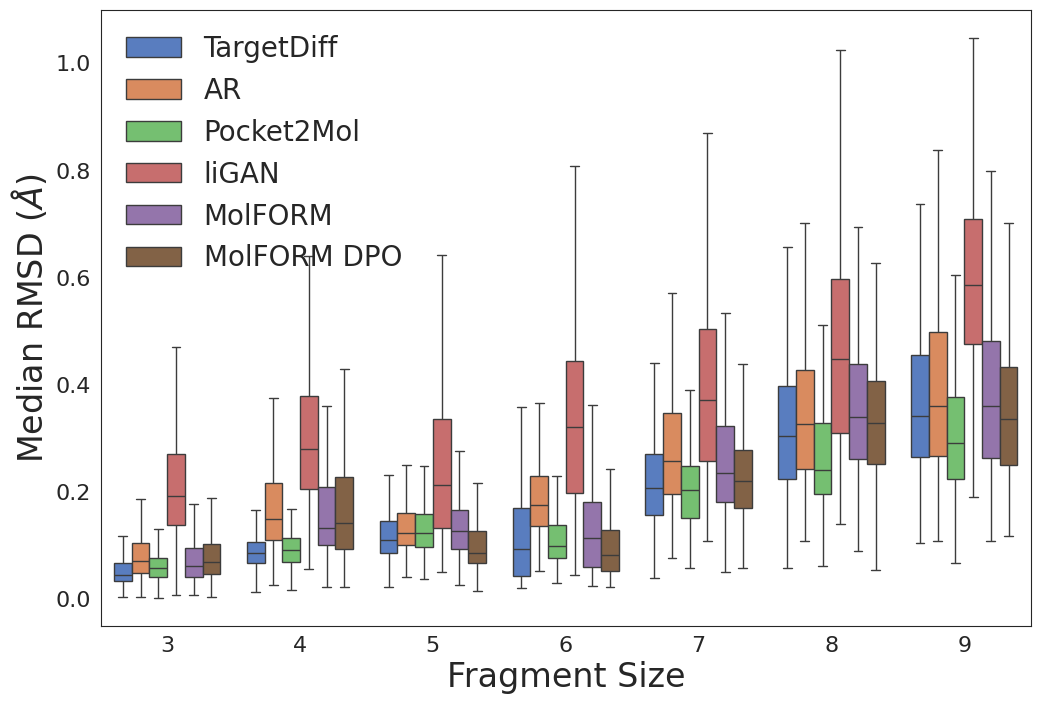}
  \caption{Median RMSD for rigid fragment before and after the force-field optimization.}
  \label{fig:RMSD}
\end{figure}

\paragraph{Evaluation} 
We collect all generated molecules across 100 test proteins and evaluate generated ligands from three aspects: \textbf{binding affinity}, \textbf{molecular properties} and \textbf{molecular structures}. For target binding affinity and molecular properties, we present our results under the best setting as \textbf{MolFORM}. Following previous work, we utilize AutoDock Vina~\citep{eberhardt2021autodock} for binding affinity estimation.  For \textbf{binding affinity}, we report the \textbf{Vina Score}, which evaluates the initially generated binding pose; the \textbf{Vina Min}, obtained after local energy minimization; and the \textbf{Vina Dock}, representing the lowest energy score from a global re-docking procedure using grid-based search. 
For \textbf{molecular properties}, we primarily report \textbf{QED} (Quantitative Estimate of Drug-likeness) and \textbf{SA} (Synthetic Accessibility).
These results are summarized in Table \ref{tab:combined_metrics}. 
For \textbf{molecular structures}, we first evaluate several geometry-related properties following the setup in ~\citet{lin2024cbgbench}. These include (i) \textbf{JSD$_{\mathrm{BL}}$} and (ii) \textbf{JSD$_{\mathrm{BA}}$}, which measure the divergence in bond length and bond angle distributions between generated and reference molecules, reflecting structural realism. (iii) \textbf{Ratio$_{\mathrm{cca}}$} denotes the proportion of atoms with steric clashes—defined as van der Waals overlaps $\geq 0.4$\AA—with protein atoms. (iv)  \textbf{Ratio$_{\mathrm{cm}}$} captures the fraction of generated molecules that contain any such clashes. These results are also included in Table \ref{tab:combined_metrics}. (v) Root Mean Square Deviation (\textbf{RMSD}) is a standard metric to evaluate the structural deviation between two sets of atomic coordinates. Lower RMSD indicates higher structural consistency reflects the extent to which rigid fragments preserve their geometric integrity before and after force field optimization, which is shown in Figure \ref{fig:RMSD}. Besides, we report the ring portion for chemical assessment in Table \ref{tab:ringsizefreq}.
For substructure-level evaluation, we analyze the Jensen-Shannon divergence (JSD) between the bond length distributions of reference and generated molecules, as shown in Table \ref{tab:jsd_bl}. Here, we mainly compare with TargetDiff, which is highly similar to our proposed MolFORM.

\begin{table}[!t]
\centering
\resizebox{0.48\textwidth}{!}{
\begin{tabular}{c|cccccc}
\toprule
Method & 3 & 4 & 5 & 6 & 7 & 8 \\
\midrule
Ref.        & 0.0130 & 0.0020 & 0.2855 & 0.6894 & 0.0098 & 0.0003 \\
LiGan       & 0.2238 & 0.0698 & 0.2599 & 0.4049 & 0.0171 & 0.0096 \\
3DSBDD      & 0.2970 & 0.0007 & 0.1538 & 0.5114 & 0.0181 & 0.0116 \\
GraphBP     & 0.0000 & 0.2429 & 0.1922 & 0.1765 & 0.1533 & 0.1113 \\
Pocket2Mol  & 0.0000 & 0.1585 & 0.1822 & 0.4373 & 0.1410 & 0.0478 \\
TargetDiff  & 0.0000 & 0.0188 & 0.2856 & 0.4918 & 0.1209 & 0.0298 \\
DiffSBDD    & 0.2842 & 0.0330 & 0.2818 & 0.2854 & 0.0718 & 0.0193 \\
DiffBP      & 0.0000 & 0.2195 & 0.2371 & 0.2215 & 0.1417 & 0.0707 \\
FLAG        & 0.0000 & 0.0682 & 0.2716 & 0.5228 & 0.0996 & 0.0231 \\
D3FG        & 0.0000 & 0.0201 & 0.2477 & 0.5966 & 0.0756 & 0.0283 \\
DecompDiff  & 0.0302 & 0.0378 & 0.3407 & 0.4386 & 0.1137 & 0.0196 \\
MolCraft    & 0.0000 & 0.0022 & 0.2494 & 0.6822 & 0.0489 & 0.0072 \\
VoxBind     & 0.0000 & 0.0062 & 0.2042 & 0.7566 & 0.0232 & 0.0021 \\
MolFORM     & 0.0000 & 0.0291 & 0.2846 & 0.4211 & 0.2033 & 0.0455 \\
\bottomrule
\end{tabular}
}
\caption{Distribution of different ring sizes across various methods.}
\label{tab:ringsizefreq}
\end{table}

\paragraph{Result analysis}
The visualization comparison result is shown in Figure \ref{fig:visul}. As shown in Table \ref{tab:combined_metrics}, we observed that MolFORM achieved comparable performance to our baseline model TargetDiff in binding affinity and molecular properties. Compared to Targetdiff, the model shows better performance in terms of Clash and Vina Dock. 
After further DPO fine-tuning, our model achieves state-of-the-art performance across nearly all metrics.

Although we use only binding affinity as the reward, DPO aligns the generation distribution toward higher-quality molecules while preserving the chemical priors from the pretrained model, thus avoiding overfitting to a single objective. As a result, metrics such as QED and SA can also improve. This is further supported by the enhanced quality of generated molecular substructures in Figure \ref{fig:dist} and Figure \ref{fig:RMSD}. Additionally, we observe that the diversity of generated molecules remains at 0.77 after DPO fine-tuning, indicating that our approach successfully avoids mode collapse while improving generation quality.

We re-trained a version of Targetdiff and applied DPO-based fine-tuning on the model. We observed that applying DPO to Targetdiff led to improvements of 2\%, 2\%, 2\% and 3\% in QED, SA, Vina score, and Vina min, while MolFORM achieved improvements of 4\%, 8\%, 14\% and 12\% on the same metrics in Table \ref{tab:vina}. This suggests that our model has greater potential to benefit from DPO fine-tuning. MultiFlow explicitly separates the generation of discrete (e.g., atom types) and continuous (e.g., 3D coordinates) modalities via dedicated flow-based branches. This factorized structure enables DPO to assign fine-grained preference signals to each modality during optimization. As DPO compares generation quality between molecule pairs, the modular design of MultiFlow allows more targeted updates—for example, refining atom types without perturbing geometry, or vice versa. This structural disentanglement enhances the model’s responsiveness to reward signals, reduces gradient interference, and ultimately makes preference optimization more effective and stable.

\begin{table}[H]
\centering

\label{tab:finetune_comparison}
\scalebox{0.68}{
\begin{tabular}{l|c|c|c|c}
\toprule
\textbf{Metric} & \textbf{TargetDiff} & \textbf{TargetDiff-DPO} & \textbf{MolFORM} & \textbf{MolFORM-DPO} \\
\midrule
Vina Score ($\downarrow$) & -5.47 & -5.58 & -5.42 & -6.16 \\
Vina Min  ($\downarrow$)  & -6.39 & -6.59 & -6.42 & -7.18 \\
QED   ($\uparrow$)      & 0.46  & 0.47  & 0.48 & 0.50 \\
SA    ($\uparrow$)      & 0.60  & 0.61  & 0.60 & 0.65 \\
\bottomrule
\end{tabular}
}
\label{fig:}
\caption{Comparison of model performance before and after DPO fine-tuning. The reported numbers are average values here. 
The results of vanilla TargetDiff are obtained by retraining the model.
}
\label{tab:vina}
\end{table}

\paragraph{Chamfer loss Impact}
We conducted ablation studies on the Chamfer DPO loss component in Table \ref{tab:chamfer_ablation}. Our experiments demonstrate that incorporating the Chamfer distance into the DPO objective significantly enhances the model's ability to preserve molecular geometric fidelity.

\begin{table}[!t]
\centering
\begin{tabular}{l|cc}
\toprule
\textbf{Metric} & \textbf{Chamfer\-w} & \textbf{Chamfer\-wo} \\
\midrule
QED   ($\uparrow$)      & \textbf{0.48}  & 0.41 \\
SA    ($\uparrow$)      & \textbf{0.60}  & 0.59 \\
Vina Score ($\downarrow$) & \textbf{-5.42} & -4.67 \\
Vina Min ($\downarrow$)   & \textbf{-6.42} & -5.76 \\
    JSD$_{\mathrm{BL}}$ ($\downarrow$)& \textbf{0.3225} & 0.4546 \\
JSD$_{\mathrm{BA}}$ ($\downarrow$)& \textbf{0.5535} & 0.6173 \\
\bottomrule
\end{tabular}
\caption{Ablation study: Comparison of model performance with (Chamfer\-w) and without (Chamfer\-wo) Chamfer loss in Vanilla Multi-model flow matching.}
\label{tab:chamfer_ablation}
\end{table}

\paragraph{Sampling efficiency}
We achieve SOTA sampling performance. While it takes on average 3428s and 6189s for TargetDiff and DecompDiff to generate 100 samples respectively, our model only uses 69s. Thanks to the efficiency of flow matching, our sampling requires only 100 steps, compared to the 1000 steps typically used in diffusion models.

\paragraph{Training details}
The model is trained via the gradient descent method Adam Kingma and Ba \cite{kingma2014adam} with init learning $\text{rate} = 5 \times 10^{-5}$, betas=(0.95, 0.999), batch size=16, and clip gradient norm=8. During training, we apply data augmentation by adding small Gaussian noise (standard deviation of 0.1) to the protein atom coordinates. Additionally, we use an exponential learning rate decay with a decay factor of 0.6 and a minimum threshold set at \(1 \times 10^{-8}\). For the DPO model training, a smaller learning rate was applied as $\text{rate} = 5 \times 10^{-8}$ with a decay factor of 0.6 and a minimum threshold set at \(1 \times 10^{-11}\) and the beta dpo parameter was set as 5.

We trained our model on two NVIDIA GH200 GPUs. The base model training converged within 12 hours and 100k steps. Moreover, the DPO model training utilized 4 hours and 20k steps.

\section{Future Work}
Our model builds upon ~\citet{guan20233d} and achieves comparable or even superior performance across multiple evaluation metrics, indicating that the proposed MultiFlow-based modeling is a promising direction for further exploration. Beyond DPO, other fine-tuning strategies and multi-objective optimization techniques also hold promise for further advancing the performance of our current model. Moreover, we employ uniform corruption for discrete flow matching, as masked variants unexpectedly underperformed in our molecular generation task—contrasting with their success in other scientific domains. Similarly, SDE sampling via Euler-Maruyama yielded inferior results compared to ODE integration, though the specific hyperparameters for stochastic sampling may require further tuning. These findings highlight the importance of tailoring both corruption strategies and sampling algorithms to the specific requirements of SBDD task.

\section{Conclusions} 
In this work, we propose to use a novel Multi flow based generation framework \textbf{MolFORM} for the protein-specific molecule generation. Extensive experiments on the CrossDocked2020 benchmark
demonstrate the strong performance of MolFORM. Our framework can be further combined with the Direct Preference Optimization (DPO) process, achieving state-of-the-art performance across almost all evaluation metrics.
Measurements on molecular substructures indicate that the improvement is not merely due to overfitting on specific metrics, but reflects a holistic enhancement of generation quality. Moreover, our comparison with TargetDiff demonstrates that this DPO-based approach offers greater potential than unimodal fine-tuning.


\section*{Acknowledgements}

This work is supported by the NSF OAC-2340011. We would like to thank
the reviewers for taking the time and effort necessary to
review the manuscript. We sincerely appreciate Dr. Chengyue Gong for valuable
comments. 

\section*{Impact Statement}

We introduce a multi-flow DPO framework that enables preference-guided generation over both discrete atom types and continuous molecular coordinates. Beyond SBDD, the generality of our framework suggests potential extensions to broader domains of computer-aided design, including small molecule generation, material discovery, and chip layout optimization. We stress the importance of ensuring the responsible deployment of this technology and caution against its misuse in harmful or unethical contexts.
\nocite{langley00}

\bibliography{example_paper}
\bibliographystyle{icml2025}

\newpage
\appendix
\onecolumn



\end{document}